\documentstyle[11pt,epsfig]{article}
\begin{document}

\title{Entropy generation and inflation in wave collision induced
pre-big-bang cosmology}

\author{A. Feinstein$^{a,}$\footnote{{\tt wtpfexxa@lg.ehu.es}}\hspace*{0.1cm}, 
K.E. Kunze$^{b,}$\footnote{{\tt kunze@amorgos.unige.ch}}\hspace*{0.1cm} and
M.A. V\'{a}zquez-Mozo$^{c,d,}$\footnote{{\tt vazquez@wins.uva.nl,
M.Vazquez-Mozo@phys.uu.nl}}
\\
\\
{\normalsize\sl $^{a}$ Departamento de F\'{\i}sica Te\'orica, 
Universidad del Pa\'{\i}s Vasco,}\\ {\normalsize\sl  Apdo. 644, E-48080 Bilbao, Spain} \\
{\normalsize\sl $^{b}$ D\'epartement de Physique Th\'eorique, Universit\'e de Gen\`eve,} \\
{\normalsize\sl 24 Quai Ernest Ansermet, 1211 Gen\`eve 4, Switzerland }\\
{\normalsize \sl $^{c}$ Instituut voor Theoretische Fysica, Universiteit van Amsterdam,
}\\ {\normalsize\sl Valckenierstraat 65, 1018 XE Amsterdam, The Netherlands }\\
{\normalsize\sl $^{d}$ Spinoza Instituut, Universiteit Utrecht, Leuvenlaan 4,}\\
{\normalsize\sl  3584 Utrecht, The Netherlands}
}

\date{}
\maketitle

\abstract{ We study inflation and entropy generation in a recently proposed pre-big-bang
model universe produced in a collision of gravitational and dilaton waves.
It is shown that enough inflation occurs provided the incoming waves are 
sufficiently weak. We also find that entropy  in this model is dynamically 
generated as the result of the nonlinear interaction of the incoming waves,
before the universe enters the phase of dilaton driven inflation. 
In particular, we give the scaling of the entropy produced in the
collision 
 in terms of the focusing 
lengths of the incoming waves.}

\section{Introduction}

Pre-big-bang cosmology (PBB) \cite{gasp} has evolved in recent years into
one of the main trends in string cosmology. In spite of its many nice
features there is a current discussion in the literature as to what extent
the resolution of the flatness and homogeneity problems within the
PBB picture requires a certain amount of fine tuning \cite{ft}. 

In a recent paper \cite{fkvm} we have proposed a cosmological scenario in
which the universe starts in a trivial state characterised by a bath of
plane gravitational and dilatonic waves which, upon collision, generate
PBB bubble universes. These bubbles act as possible seeds for
a Friedman-Robertson-Walker (FRW) universe in the spirit of PBB 
cosmology. The proposed scenario \cite{fkvm} should be seen as a realisation of the picture
of Bounanno, Damour and Veneziano (BDV) \cite{bdv}.

The aim of this Letter is to study some relevant phenomenological implications
of the wave collision induced PBB cosmology.
To this end we will be  focusing our attention on
two issues: the conditions for successful inflation and the production of entropy
in the scenario.

 Though it was shown in \cite{fkvm} that there exists  a
dense region in parameter space,  for which the PBB inflation takes place,
it is important to clarify further as to whether such favourable initial conditions
would lead to sufficient inflation. In particular, we will see that inflationary requirements 
leading
 to consider large black holes as the seeds of PBB 
bubbles in the original BDV picture here imply the weakness of the incoming waves.
Therefore, these waves
can be regarded as emerging from perturbations of flat space-time.
With respect to entropy production, unlike in the original BDV scenario
where the origin of entropy remains rather  vague, we will see that 
entropy in our model is generated dynamically   as a result of the wave collision. 

In the wave collision induced
model universe, one starts in the remote past with non-interacting
plane gravi-dilatonic waves. It was argued in \cite{fkvm} that such a state should
be identified as one with a  minimum entropy, for both gravitational and matter
sectors. This assumption is based on the following reasons:

\begin{itemize}
\item[i)]  Simplicity: Plane waves have no nonvanishing curvature
invariants, and thus are, apart from Minkowski spacetime, the simplest
geometries one may think of.

\item[ii)]  Symmetry requirements: With the advent of the string theory and
its alphabetic generalisations (M-theory and F-theory) our view on the
initial state of the universe has changed considerably. In particular one is
led to think that the primordial universe should be described by some exact
string background which at later time evolves into a FRW-like universe. In
this light, Penrose's proposal \cite{penrose} of assigning zero gravitational
entropy to backgrounds with vanishing Weyl curvature (a particular case of
which is the FRW metric) seems to be too restrictive. This is especially
evident in the context of PBB cosmology where the universe originates in a
highly perturbative state and only evolves into a FRW phase after a graceful
exit phase. Plane gravitational waves, being exact string backgrounds \cite
{horstief}, are ideal candidates to represent this PBB primordial
 state \cite{gr}.

\item[iii)]  Absence of particle creation: The fact that in the vicinity of
plane waves there is no quantum creation of particles enforces the idea that
plane waves are zero gravitational entropy states. Even using Penrose's
original suggestion \cite{penrose,no} where the ``number of gravitons''
contained in the gravitational field could be adopted as a measure of
entropy, plane waves are likely to be associated with minimum gravitational
entropy.
\end{itemize}

\begin{figure}[tbp]
\centerline{\epsfig{file=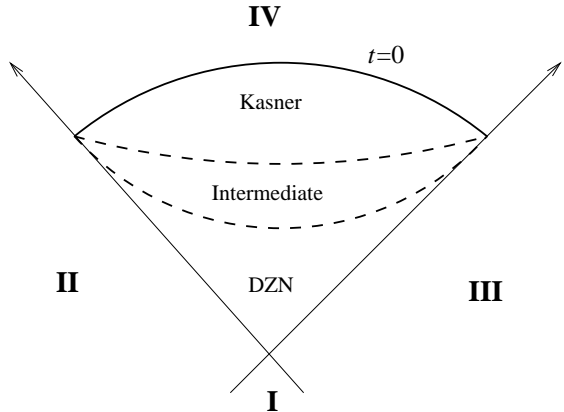, width=2.9in}}
\caption{Interaction region in wave collision induced PBB cosmology.}
\end{figure}

We now briefly summarise some features of the model universe described in 
\cite{fkvm} (the reader is referred for details to the original paper). The
universe starts in a distant past with weak, strictly {\it plane} waves
propagating on a flat background. Eventually these waves collide breaking the
plane symmetry in the interaction region (Fig. 1). Just after the collision takes place
at some $t_{i}<0$, 
physics is dominated by the superposition of two noninteracting null fluids
so the geometry is described by a Doroshkevich-Zeldovich-Novikov (DZN) line element
\cite{dzn}. This phase is followed by an intermediate region where the nonlinearities
of gravity become important and the evolution then becomes nonadiabatic.
Finally, as we approach the caustic singularity the universe enters a Kasner phase
(for $t_{K}<t<0$) where matter is effectively described in terms of a stiff fluid 
and the evolution is adiabatic. In this region the geometry is given by a 
generalised Kasner line element whose exponents depend on one spatial coordinate 
alone. In Einstein frame this metric can be written as 
\begin{eqnarray}
ds^{2}=-dt^{2}+\left({-t\over -t_{K}}\right)^{2\alpha _{1}(z)}dx^{2}
+\left({-t\over -t_{K}}\right)^{2\alpha
_{2}(z)}dy^{2}+\left({-t\over -t_{K}}\right)^{2\alpha _{3}(z)}dz^{2},
\label{met}
\end{eqnarray}
where we have normalized the scale factors to their values at the beginning
of the Kasner epoch, $t=t_{K}$. 

The generalised Kasner exponents $\alpha_{i}(z)$ are determined by the initial data on the
null boundaries of the interaction region. In terms of the gravitational and
scalar source functions $\epsilon (z)$ and $\varphi (z)$ 
given, in turn, by quite complicated line integrals of incoming data \cite{fkvm}, the
Kasner exponents can be written as 
\begin{eqnarray}
\alpha _{1}(z) &=&{\frac{1+\epsilon (z)}{a(z)+2}},  \nonumber \\
\alpha _{2}(z) &=&{\frac{1-\epsilon (z)}{a(z)+2}},  \nonumber \\
\alpha _{3}(z) &=&{\frac{a(z)}{a(z)+2}},  \label{ke}
\end{eqnarray}
where $a(z)\equiv {\frac{1}{2}}[\epsilon (z)^{2}+\varphi (z)^{2}-1]$. They satisfy
the following relations 
\begin{eqnarray}
\sum_{i=1}^{3}\alpha _{i}(z)=1,\hspace{1cm}\sum_{i=1}^{3}\alpha
_{i}(z)^{2}=1-{\frac{2\varphi (z)^{2}}{[a(z)+2]^{2}}}.
\label{kas}
\end{eqnarray}
On the other hand the dilaton field near the singularity behaves like 
\begin{eqnarray}
\phi (t,z)=\beta (z)\log {\left( \frac{-t}{-t_{K}}\right) }+\phi _{0}
\label{dil}
\end{eqnarray}
with 
\begin{eqnarray}
\beta (z)={\frac{2\varphi (z)}{a(z)+2}}
\label{beta}
\end{eqnarray}
and $\phi _{0}$ being some constant zero mode that determines the string coupling
constant at the beginning of dilaton driven inflation (DDI), 
$g_{st}=e^{{1\over 2}\phi_{0}}$. Since we have normalized the metric and the dilaton at the
beginning of DDI the natural scale in Einstein frame in this regime will be
the Planck scale at $t_{K}$, $\ell_{Pl}=e^{{1\over 2}\phi_{0}}\ell_{st}$.  
We will assume in the following that 
the dilaton field does not vary very much from the initial boundaries to the
onset of the Kasner regime, so $\phi_{0}\sim \phi(t\rightarrow -\infty)$. 
In this case asymptotic past triviality implies that $\phi_{0}\ll -1$.

\section{Successful inflation}

In \cite {fkvm} it was shown that there exists a dense set of points in 
the $\epsilon(z)$-$\varphi (z)$ plane for which the models undergo PBB 
inflation in string frame. These inflationary models correspond in Einstein
frame to a contracting universe where all Kasner exponents $\alpha_{i}(z)$
are positive. 

The question however remains of whether successful inflation, i.e. enough number
of $e$-foldings, can be achieved in the collision induced PBB scenario and if so
how the amount of inflation relates with the initial conditions in the collision. In
this section we will analyse how the requirement on the number of $e$-foldings
translates into the strengths of the incoming waves. Since  
the Kasner exponents are analytically related to the 
initial conditions, we can perform a fully general analysis of
the PBB inflationary era by just looking at the Kasner regime. 
Different particular solutions in the interaction region will be encoded into different
source functions $\epsilon(z)$ and $\varphi(z)$, which in turn will translate into
different set of Kasner exponents through Eqs. (\ref{ke}). 

In the picture of colliding waves one starts with the collision of two
gravi-dilatonic waves with some focal lengths $L_{1}$ and $L_{2}$. 
As a result of their interaction, the waves will focus each other to form a
caustic singularity in the interaction region after a time 
$|t_{i}|\sim \sqrt{L_{1}L_{2}}$ has elapsed, as measured in the Einstein frame 
\cite{szek}. Physically speaking, the focusing lengths of the incoming waves
give the measure of their strength; long focal lengths correspond to weak waves
and thus it takes very long  time till a singularity forms, whereas strong waves are 
characterized by short focal lengths.

The amount of inflation occuring during some time period in the string 
frame in each direction can
be expressed  as the ratio of the initial ($\tilde{t}=\tilde{t}_{1}$) and 
final ($\tilde{t}=\tilde{t}_{2}$)
comoving Hubble radii given by the three functions 
\begin{eqnarray}
Z_{i}={\frac{\tilde{a}_{i}(\tilde{t}_{2})\tilde{H}_{i}(\tilde{t}_{2})}{
\tilde{a}_{i}(\tilde{t}_{1})\tilde{H}_{i}(\tilde{t}_{1})}},\hspace{1cm}
(i=1,2,3),
\label{Z}
\end{eqnarray}
where the tilde indicates quantities in the string frame.
Evaluating explicitly $Z_{i}$ and using Einstein frame variables 
in the Kasner regime where both the
expressions of $a_{i}(t)$ and $\phi (t)$ are given by Eqs. (\ref{met}) and (\ref{dil}), 
we find 
\begin{eqnarray}
Z_{i}=\left( \frac{-t_{2}}{-t_{1}}\right) ^{\alpha _{i}-1}.
\end{eqnarray}
Notice that the largest $Z_{i}$ corresponds (for $t_{1}<t_{2}$) to the
direction with the smallest Einstein frame Kasner exponent.

In order to analyse the total amount of inflation during DDI we still need to 
determine when the inflationary regime comes to an end.
We  consider that
DDI begins at $t_{1}=t_{K}$ when the universe enters the Kasner regime. On the
other hand, to determine the moment at which inflation is over
we should look at the instant of time when the low energy approximation breaks down.
Thus, we will take $t_{2}=t_{f}$ as the time at which either curvature reaches the
string scale or  the effective string coupling becomes of order 1. 

Let us work out the first possibility. By imposing ${\tilde{H}}_{i}^{2}(\tilde{t})\sim 
\ell _{st}^{-2}$ in at least one of the three directions, we find for $t_{2}$ the 
following expression in the Einstein frame 
\begin{eqnarray}
[t_{f}]_{\alpha'}\sim |p_{\rm min}|^{\frac{2}{\beta +2}}
\left({\ell_{st}\over -t_{K}}\right)^{2\over \beta+2}t_{K}
=\left| {\frac{2\alpha
_{\rm min}+\beta }{\beta +2}}\right| ^{\frac{2}{\beta +2}}
\left({\ell_{st}\over -t_{K}}\right)^{2\over \beta+2}t_{K},
\label{curv}
\end{eqnarray}
where $p_{\rm min}$ and $\alpha_{\rm min}$ are the smallest Kasner exponents
in string and Einstein frame respectively. The reason for this is that in the 
inflationary region $2/(\beta +2)>1$ \cite{fkvm}, therefore we conclude that in
order to define $t_{f}$ we should take the largest $|p_{j}|$. Since in 
this region all $p_{i}$ are negative the direction 
with the largest value of $|p_{i}|$ corresponds to the one with the smallest
Kasner exponent (in both the Einstein and the string frame).
 
Yet a different constraint on $t_{f}$ can be obtained by looking at the time
when the string theory coupling becomes of order 1. 
Given the form of the dilaton field in
the Kasner regime (\ref{dil}) we have that the effective string coupling is
given by
\begin{eqnarray}
g_{{\rm eff}}=\left( {\frac{-t}{-t_{K}}}\right) ^{{\frac{1}{2}}\beta }e^{
{\frac{1}{2}}\phi _{0}}.
\label{scoup}
\end{eqnarray}
In particular, it can be seen that, since $\beta
<0$ for the solutions leading to PBB inflation, the effective coupling blows
up when $t\sim 0$. If we take $t_{f}$ as the time at the onset of strong
coupling effects ($g_{{\rm eff}}\sim 1$) where our approximation breaks down, we find 
$$
[t_{f}]_{g_{st}}\sim e^{-{\frac{\phi _{0}}{\beta }}}t_{K}.
$$

By looking at Eqs. (\ref{curv}) and (\ref{scoup}) we see that long periods of
inflation ($t_{\rm inf}=|t_{K}-t_{f}|$) can be achieved by either taking 
$|t_{K}|\gg \ell_{st}$ or $e^{\phi_{0}}\ll 1$. The first case will correspond, generically,
to the case of long focusing times (i.e. very weak waves), whereas the second
condition is just equivalent to asymptotic past triviality \cite{bdv}.

To quantify the 
amount of inflation we can now proceed to evaluate $Z_{i}$ using the 
definition (\ref{Z}) to find (cf. \cite{cl})
\begin{eqnarray}
Z_{i}={\rm Min}\,\left[ |p_{\rm min}|^{\frac{2(\alpha _{i}-1)}{\beta +2}}
\left({-t_{K}\over \ell_{st}}\right)^{2(1-\alpha_{i})\over \beta+2}
,e^{{\frac{1-\alpha _{i}}{
\beta }}\phi _{0}}\right],
\label{zi}
\end{eqnarray}
and in order to solve the flatness/horizon problem we have to require 
\begin{eqnarray}
Z_{i}\stackrel{>}{\sim }e^{60}.
\label{ir}
\end{eqnarray}
Now, the condition for getting a sufficiently large value for $Z_{i}$ translates
into a condition on two parameters, 
namely $t_{K}$ (which depends on the initial focal lengths 
$L_{1}$, $L_{2}$) and the zero mode of the dilaton $\phi _{0}$. 

In any case, it is possible to find which kind of corrections ($\alpha'$ or $g_{st}$)
will end inflation by looking at Eqs. (\ref{curv}) and (\ref{scoup}). From there
we can get a critical value of $t_{K}$ such that when
$$
t_{K} > t_{\rm crit}\equiv -|p_{\rm min}|e^{\phi_{0}\over \beta}\ell_{Pl}
$$
DDI will be terminated by the onset of $\alpha'$ corrections. On the other hand
if $t_{K}<t_{\rm crit}$ string loop corrections will become relevant first and 
will signal the end of the inflationary period. Notice that the requirement of
asymptotic past triviality, together with the fact that for inflationary solutions
$\beta<0$, implies for generic models that $|t_{\rm crit}|\gg \ell_{st}
\gg\ell_{Pl}$ so the crossover between the two regimes occurs well inside 
the region of applicability of the low energy approximation.

We can now translate the condition (\ref{ir}) into constraints on the parameters
of the initial waves (their focal lengths $L_{1}$ and $L_{2}$) and the
zero-mode of the dilaton field $\phi_{0}$. In the situation when the duration of
inflation is determined by the excitations of massive string states ($t_{K}>t_{\rm crit}$)
$Z_{i}$ is given by the first entry on the right hand side of Eq. (\ref{zi}).
Thus, the requirement (\ref{ir}) implies for a model with  reasonably small deviations 
from isotropy, i.e. $|p_{\rm min}|^{2(\alpha_{i}-1)\over\beta+2}\sim 1$, 
$$
t_{i}<t_{K}< -e^{{30(\beta+2)\over 1-\alpha_{\rm min}}}\ell_{st}.
$$
Since $|t_{i}|\sim \sqrt{L_{1}L_{2}}$ we have a bound on the strength of 
the incoming gravi-dilatonic waves
\begin{eqnarray}
L_{1} L_{2} \stackrel{>}{\sim} e^{{60(\beta+2) \over 1-\alpha_{\rm min}}}\ell_{st}^{2}.
\label{bnd}
\end{eqnarray}
Therefore, assuming for simplicity that both incoming waves  have similar focal
lengths, we arrive at the conclusion that successful inflation will be achieved for
extremely weak waves.  
In the cases with $t_{K}<t_{\rm crit}$ the end of inflation is triggered
by 
string loops corrections we get a bound on the zero mode of the dilaton, namely
\begin{eqnarray}
\phi_{0}\stackrel{<}{\sim}{60\beta\over 1-\alpha_{\rm min}}\sim -10^{2}
\label{dbnd}
\end{eqnarray}
which corresponds to a very weakly coupled string theory at the beginning of DDI.
Actually, this bound of the dilaton field leads to the same constraint on the focal
lengths of the incoming waves found in Eq. (\ref{bnd}). This can be realized by 
combining Eq. (\ref{dbnd}) with the condition that whenever inflation is terminated
by string loop corrections $|t_{K}|>|t_{\rm crit}|$. It is remarkable that, 
independently of the physical mechanism responsible for the end of the inflationary 
regime, one obtains the same bound on the focal lengths of the colliding waves.

Consequently, we conclude that successful inflation occurs for very
weak incoming waves and/or very weak initial string coupling. The condition of
having extremely weak initial waves is very satisfactory, since these
 may be considered as emerging from
 fluctuations in flat space-time. Besides, since the focusing time for these
waves will be very long we can argue that, on general grounds, for sufficiently 
weak waves $t_{K}<t_{\rm crit}$ and inflation therefore  will be terminated by the onset
of string loop corrections. If we compare with the original BDV proposal we see that
our condition on the strength of the incoming waves corresponds to their 
condition of starting with a large black hole. It is worth noticing, however, that
in the wave collision induced version of the BDV scenario the ``fine tuning'' 
of the size of the black hole is automatically achieved if the incoming waves 
are viewed as small fluctuations on an otherwise trivial background. 

Incidently, we find as a bonus that particle production in the interaction region
for weak incoming waves is negligible \cite{dv,temp}. Thus, as in  
\cite{bdv}, emerging homogeneity is not expected to be spoiled by 
quantum corrections.

\section{Entropy production}

A different phenomenological issue which will be addressed 
 here is the entropy production. We  start with
the assumption that before the  collision between the waves takes place, both 
matter and  gravitational entropy are zero \cite{fkvm}. As the result of the 
collision, entropy will be generated
from its zero value close to the null boundaries to a 
nontrivial entropy content as the universe approaches the singularity. 
Since the rate at which gravitational entropy grows is difficult to estimate 
we will focus our attention on the production of matter entropy.

As explained in the Introduction, the interaction region of the two incoming waves
can be divided into three regions (Fig. 1). Just after the collision takes place at
$t_{i}$ the matter content of the universe can be satisfactorily described in terms of a 
superposition of the two incoming  non-interacting null fluids and
 the metric is given by a DZN line
element. In this regime the evolution is very approximately adiabatic and almost no
entropy is produced. This almost linear regime comes to an end
 as soon as the gravitational nonlinearities
 take over and we enter an intermediate phase where the dynamics of the
universe is dominated by both velocities and spatial gradients. Now the evolution 
is no longer adiabatic and matter entropy is generated. In this regime the matter 
content of the universe cannot be
described in terms of a perfect fluid equation of state and some effective macroscopic 
description of the fluid, such as an anisotropic fluid or some other phenomenological 
stress-energy tensor, should be invoked \cite{taub}. 

The production of entropy
will stop at the moment the velocities begin to dominate over spatial gradients and the 
evolution becomes  adiabatic again, the matter now being described by a perfect
fluid with stiff equation of state $p=\rho$. The transition to an adiabatic phase
will happen either before or at $t_{K}$ when the universe enters  the Kasner phase. 
{}From that moment on no further entropy is produced up to the end of DDI.
The relative duration of these three regimes crucially depends
on the strength of the incoming waves and on the initial data.
It is straightforward to show that for both 
gravitational and scalar waves the ratio of spatial gradients versus
time derivatives dies off as $1/\sqrt{L_1L_2}$  and therefore it takes the 
universe a time of order $\sqrt{L_1L_2}$ before the Kasner-like regime is reached, 
 fixing the  duration of the  nonadiabatic phase.

As discussed above, all the 
entropy is generated in the intermediate phase between the adiabatic DZN and 
Kasner regimes. On the other hand, in the Kasner regime the total entropy generated
in this intermediate region is carried adiabatically by the perfect stiff fluid 
represented by the dilaton field $\phi(t)$. If we consider a generic bariotropic 
equation of state, $p=(\gamma -1)\rho $, the first law of thermodynamics implies 
that the energy density $\rho$ and entropy density $s\equiv S/V$ evolve with the
temperature $T$ as 
\begin{eqnarray}
\rho &=&\sigma \,T^{\frac{\gamma }{\gamma -1}}  \label{rho} \\
s &=&\gamma \,\sigma \,T^{\frac{1}{\gamma -1}},  \label{s}
\end{eqnarray}
where the temperature is
given by the usual relation $T^{-1}=(\partial s/\partial \rho )_{V}$.
Here $\sigma $ is a dimensionful constant which in the particular 
case of radiation ($\gamma=4/3$) is just the 
Stefan-Boltzmann constant. Using Eqs. (\ref{rho}) and (\ref{s}) we easily
find that 
for a stiff perfect fluid ($\gamma =2$) the entropy density $s$ can be expressed in terms of
the energy density as 
\begin{eqnarray}
s=2\sqrt{\sigma\rho}. 
\label{srho}
\end{eqnarray}
It is important to stress here that $\sigma$ is a parameter that
gives the entropy content of, and thus the number of degrees of freedom that
we associate with, the effective perfect fluid. In principle it
could be computed provided a microscopical description of the fluid is available. 
However in the case at hand, the stiff perfect fluid is just an effective description of
the classical dilaton field in the Kasner regime. Since the evolution in the DDI phase is
adiabatic, $\sigma$ can be seen as a phenomenological
parameter that measures the amount of entropy generated during the intermediate 
region where the dilaton field should be described by an imperfect effective
fluid. This effective description of the dilaton condensate, and the
entropy generated,  would then depend on a number of phenomenological parameters.

We can now give an expression for the total entropy inside a Hubble volume 
at the beginning of DDI in terms of the initial data. The energy density carried by
the dilaton field in the Kasner epoch is given by 
$$
\rho(t)={\dot{\phi}^2\over 4\ell_{Pl}^2}={\beta^2\over 4\ell_{Pl}^2 t^2},
$$
where $\beta$ is the source function for the dilaton defined in Eq. (\ref{beta}).
From here we get the entropy density by applying Eq. (\ref{rho}). Thus, the
total entropy inside a Hubble volume at the beginning of the Kasner regime 
can be written in terms of the Kasner exponents and the source function for the dilaton as
$$
S_{H}(t_{K})={t_{K}^{2}\over \ell_{Pl}^2}\sqrt{\sigma\ell_{Pl}^2}
{|\beta|\over \alpha_{1}\alpha_{2}\alpha_{3}}
$$
Now we can write $t_{K}=\eta \sqrt{L_{1}L_{2}}$ with 
$0<\eta\stackrel{<}{\sim} 1$, so finally
we arrive at
$$
S_{H}(t_{K})=\left[\eta^2\sqrt{\sigma\ell_{Pl}^2}{|\beta|\over \alpha_{1}\alpha_{2}\alpha_{3}}
\right]{L_{1}L_{2}\over \ell_{Pl}^2} 
\equiv \kappa {L_{1}L_{2}\over \ell_{Pl}^2}.
$$

It is interesting to notice that this scaling for the entropy in the
Hubble volume at 
the beginning of DDI can be also retrieved by considering a particular example when,
as a result of the wave collision,
a space-time locally isometric to that of a Schwarzschild black hole is produced
in the interaction region. In that case
the focal lengths of the incoming waves are related with the mass of the black hole
by $M=\sqrt{L_{1}L_{2}}/\ell_{Pl}^{2}$ (see for example \cite{dv}). Now we can write
the Bekenstein-Hawking entropy of the black hole $S=4\pi\ell_{Pl}^{2}M^{2}$ in terms of
the focal lengths of the incoming waves as $S=4\pi L_{1}L_{2}/\ell_{Pl}^{2}$. 
This analogy is further supported by the fact that the temperature of the created 
quantum particles in both the black hole and the 
colliding wave space-time scales like $T\sim 1/M$ and $T\sim 1/\sqrt{L_{1}L_{2}}$
respectively \cite{dv,temp}, as well as by the similarities between 
the thermodynamics of black holes and stiff fluids \cite{zurek}.
It is important to notice that 
this scaling of the temperature of the created particles    
with the focal lengths implies that, whenever enough inflation occurs,   
the contribution of these particles to the total entropy is negligible.

One may have thought, in principle,  that it is possible to avoid the
entropy production before the DDI by just taking a solution in the
interaction region for which the evolution is globally adiabatic. The
simplest possibility for such a solution would be a Bianchi I metric for
which the Kasner regime extends all the way back to the null boundaries.
This, however, should be immediatly discarded due to the constraints 
posed by the boundary conditions in the colliding wave problem \cite{grif}.
Or, put in terms of the null data, it is not possible to choose the
initial data on the null boundaries in such a way that the metric is
globally of Bianchi I type in the whole interaction region and at the same
time ${\cal C}^1$ and piecewise ${\cal C}^2$ across the boundary.

Before closing this Section we briefly discuss whether the entropy produced 
before the DDI phase complies with different cosmological entropy bounds. 
To study Bekenstein's entropy bound \cite{bekb} it is convenient to define the
function \cite{ms} ($\hbar=c=1$)
\begin{eqnarray}
f\equiv \frac{1 }{2\pi }\frac{s}{\rho R}\leq 1,
\label{ff}
\end{eqnarray}
where $R$, the effective radius of the system, can be defined following Refs. \cite{bek,ms}
as the maximal extension of the particle horizon in the three spatial directions. 
If we
concentrate our attention on the region near the end of inflation $t\sim
t_{f}$ where we are in the Kasner regime it is
easy to see that the particle horizon is largest in the direction with the
smallest Kasner exponent $\alpha_{\rm min}$. Then, the function $f^{(P)}$ scales with time as
(the superscript indicates that we are using the particle horizon to define the size of
the system)
\begin{eqnarray}
f^{(P)}(t)\sim 
\left( \frac{-t}{-t_{K}}
\right) ^{1-\alpha_{\rm min}}.
\label{fop}
\end{eqnarray}
This is of course expressed in the Einstein frame. As it turns out, 
evaluating Bekenstein bound in the string frame just comes down
to substituting Einstein frame quantities by string frame quantities in the 
above equation.

Looking at Eq. (\ref{fop}) we see that, since $1-\alpha_{\rm min}>0$ the
function $f^{(P)}(t)$ goes to zero as $t$ approaches the singularity. Actually, the value
of this function at the end of DDI scales with the number of $e$-foldings in 
the direction of $\alpha_{\rm min}$ as $f^{(P)}(t_{f})\sim Z_{\rm min}^{-1}$. 
Therefore Bekenstein bound is well satisfied near the curvature singularity. 
On the other hand, if we were to evaluate the function $f^{(P)}$ at earlier times a
detailed analysis of the intermediate region where entropy is produced would be needed.

In the cosmological version of Bekenstein entropy bound the size of the system
is determined by the particle horizon. However, one could use instead 
the event horizon which is proportional to the Hubble radius. In the
Kasner regime the Hubble radius in the $j$-th direction is given by 
$$
|H_{j}(t)|^{-1}\equiv \left| {\frac{a_{j}(t)}{\dot{a}_{j}(t)}}
\right| ={\frac{-t}{\alpha _{j}}} 
$$
so by taking $R$ in (\ref{ff}) to be $R={\rm Max}
\,[|H_{1}|^{-1},|H_{2}|^{-1},|H_{3}|^{-1}]$ we are led again to select the
direction with the smallest $\alpha _{j}$ (recall that for inflationary models
$0<\alpha_{j}<1$). Evaluating the function $f^{({H}
)}(t)$ we find that it is constant for the whole Kasner epoch and
solely determined by the initial conditions
\begin{eqnarray}
f^{({H})}(t)\simeq {\frac{2 \sqrt{\sigma\ell_{Pl}^2}}{\pi |\beta |}}\alpha
_{\rm min}
\label{veb}
\end{eqnarray}
where by $\simeq $ we mean that we are dropping the multiplicative constant
of order one relating the event horizon with the Hubble radius. The
condition $f^{({H})}\leq 1$ corresponds then to the entropy
bound  proposed by Veneziano (see \cite{gent} and references therein).  
This bound will be satisfied depending on the value of the dimensionless
quantity $\sigma\ell_{Pl}^2$ that, as discussed above, measures the total 
amount of entropy generated during the intermediate regime.
The corresponding function
in string frame, $\tilde{f}^{({
H})}$, is also constant and can be obtained by substituting the constants on
the right hand side of (\ref{veb}) by their string frame counterparts. 

We therefore see from (\ref{veb}) that the Hubble entropy 
bound will be satisfied depending 
on initial conditions and the thermodynamics of the intermediate region.

It is interesting to notice that, since the 
Hubble radius is a ``local'' quantity
which only depends on the expansion/contraction rate at a given instant, the function $
f^{(H)}(t)$ contains no information about the history of the universe
beyond the one encoded in the value of $\sigma\ell_{Pl}^2$. 
In particular, there is
no dependence on what happened before the universe entered the Kasner regime
between the times $t_{i}$ and the time $t_{K}$ . This contrasts
with $f^{(P)}(t)$ which, due to the nonlocal character of the particle
horizon, encodes even in the Kasner regime information about the pre-Kasner
epoch. Furthermore, one should not
be surprised that Veneziano's bound is time-independent in
this example. By using the event horizon one is basically assigning a black
hole (maximum) entropy to the solution which in the case of the stiff fluid
coincides with that of the black hole anyway (cf.  \cite{zurek}).

\section{Discussion}

In this Letter we have investigated some phenomenological aspects of the
picture for PBB cosmology proposed in Ref. \cite{fkvm}. In particular, we
have focused our attention on two special aspects of the model:
the conditions for successful inflation and the entropy generation.

As to the requirement of successful inflation we have found that 
in order to meet the necessary number of $e$-foldings to solve the homogeneity
and flatness problems the focal lengths of the incoming waves have to 
be exponentially large with respect to the string scale. Actually it is 
interesting that the bound on the product $L_{1}L_{2}$ is independent of the physical 
mechanism terminating inflation, $\alpha'$ corrections or strong coupling effects.
This provides quite a natural
picture where the waves might be seen as small perturbations on
a flat space-time producing upon collision the PBB bubbles. One of the
alleged ``fine-tunings'' of the BDV picture was the necessity of starting with
very large black holes in order to produce sufficient inflation during the 
PBB phase. In our picture, this requirement is translated into a rather ``natural'' condition 
of having extremely weak incoming waves.  

With respect to the production of entropy we have seen, that entropy is generated
dynamically as the result of the nonlinear interaction of the incoming waves. 
Thus, the following picture emerges. 
As argued in \cite{fkvm} the universe starts as a bath of gravitational and
dilatonic plane waves with zero total (matter+gravitational) entropy. The
waves eventually interact and, before producing a singularity, the universe  
passes through a stage of nonadiabatic evolution where entropy is created
(Fig. 1). The process of entropy generation  ends when the
universe enters the regime of adiabatic (perfect fluid) evolution. This
happens at the onset of the Kasner phase or earlier, depending whether
velocities dominate over spatial gradients at or before entering
the Kasner regime. We have also seen that the total entropy inside a Hubble volume
at the beginning of DDI is related to the focal lengths of the incoming waves
by $S=\kappa L_{1}L_{2}/\ell_{Pl}^{2}$, with $\kappa$ a numerical constant depending
on the initial conditions of the collision and the thermodynamics of the effective
imperfect fluid describing the dilaton condensate in the nonadiabatic intermediate regime.

An immediate question to be asked is whether the analytical relation between the initial 
data and the structure of the singularity \cite{fkvm} in this scenario allows a quantitative
evaluation of the entropy produced. This, unfortunately, cannot be done since 
any analysis of this type would require some input about the
thermodynamical properties of the effective fluid in the {\it whole} interaction region.
This is easy to understand on physical grounds, since the initial data only determines the
dynamics of the background fields (the metric and the dilaton in this case), but not
the thermodynamics of the dilaton condensate. In order to specify from first principles
the phenomenological parameters required for the description of the fluid in the 
nonadiabatic regime (which would determine the amount of entropy generated and therefore
the value of $\sigma\ell_{Pl}^2$) one would need to invoke a microscopical description of 
the matter content. This situation is analog to the case of standard FRW cosmology, where
the gravitational dynamics of a radiation dominated universe does not determine the 
value of the Stefan-Boltzmann constant.

We have checked that the Bekenstein entropy bound is satisfied close to the
singularity and can only be violated at earlier times. However, the conditions 
favourable for a long lasting inflation seem to comply with the Bekenstein
entropy bound. 
On the other hand we have seen
that the Hubble entropy bound will be satisfied depending 
on the value of the Kasner exponents (and as a consequence on the initial
data on the null boundaries) as well as on the amount of entropy 
generated during the intermediate regime.

To close, the results of our analysis suggest a picture in which 
some collisions occur within a primordial ``cold'' bath of weak gravi-dilatonic
waves. Before entering into the DDI stage the universe undergoes
 a nonadiabatic transient epoch where entropy is produced. If the waves are weak enough (as
one would expect if they result from small fluctuations of the backgrond) the resulting PBB 
bubble can become the seed of a FRW universe of the kind we see at present. 
In order to study  the scenario analytically we have confined the initial data
to strictly plane incoming waves. 
Nonetheless, it would be great to understand the problem in terms of more
general $pp$ waves which have finite tranverse size. Note, however, that
unlike plane waves, $pp$ waves do polarise the vacuum, and assigning initial zero
entropy to those would be somehow less natural.

\section*{Acknowlegements}

It is a pleasure to thank Jacob Bekenstein, Enric Verdaguer and 
especially Gabriele Veneziano for 
useful discussions and correspondence. 
A.F. has been supported by University of the Basque Country Grant
UPV 122.310-EB150/98, General University Research Grant UPV172. 310-G02/99
 and Spanish Science Ministry Grant PB96-0250.
K.E.K. acknowledges the support of the Swiss National Science Foundation.
The work of M.A.V.-M. has been supported by FOM ({\it Fundamenteel Onderzoek 
van der Materie}) 
Foundation and by University of the Basque Country Grants UPV 063.310-EB187/98 and UPV
172.310-G02/99, and Spanish Science Ministry Grant AEN99-0315.
K.E.K. and M.A.V.-M. wish to thank the Department of Theoretical Physics of 
The University of the Basque Country for hospitality. M.A.V.-M. also thanks
CERN Theory Division for hospitality during the final stages of this work.

\end{document}